\documentclass[preprint]{aastex}
\usepackage{emulateapj5,apjfonts,epsf,epsfig,graphicx,natbib}

\newcommand{\mn}{{Mon.\@ Not.\@ Roy.\@ Ast.\@ Soc.\ }}
\newcommand{\asta}{{Astron.\@ Astrophys.\ }}

\newcommand{\etal}{{\it et al.~}}
\newcommand{\ie}{{i.e.}}
\newcommand{\etc}{{\it etc.}}

\newcommand{\eg}{{e.g.,~}}

\newcommand{\beq}{\begin{equation}}
\newcommand{\eeq}{\end{equation}}
\newcommand{\ber}{\begin{eqnarray}}
\newcommand{\eer}{\end{eqnarray}}
\newcommand{\lleq}{\lower0.9ex\hbox{ $\buildrel < \over \sim$} ~}
\newcommand{\ggeq}{\lower0.9ex\hbox{ $\buildrel > \over \sim$} ~}
\newcommand{\lsim}{\ \lower-1.5pt\vbox{\hbox{\rlap{$<$}\lower5.3pt\vbox{\hbox{$\sim$}}}}\ }
\newcommand{\gsim}{\ \lower-1.5pt\vbox{\hbox{\rlap{$>$}\lower5.3pt\vbox{\hbox{$\sim$}}}}\ }

\newcommand{\pr}{\prime}

\newcommand{\de}{dark energy}
\newcommand{\De}{Dark energy}
\newcommand{\DE}{Dark Energy}

\newcommand{\ld}{\Lambda}
\newcommand{\Perts}{Perturbations}
\newcommand{\perts}{perturbations}
\newcommand{\pert}{perturbation}

\newcommand{\HH}{{\cal H}}
\newcommand{\cs}{c_{s,DE}^2}
\newcommand{\ca}{c_{a,DE}^2}
\newcommand{\dlt}{\delta_{DE}}
\newcommand{\vel}{v_{DE}}
\newcommand{\ww}{w_{DE}}

\newcommand{\w}{w_0}
\newcommand{\wm}{w_m}
\newcommand{\at}{a_t}
\newcommand{\dt}{\Delta_t}

\newcommand{\omt}{\Omega_{0 \rm m}}

\newcommand{\omde}{\Omega_{DE}}

\newcommand{\om}{\Omega_{\rm m}}

\newcommand{\omk}{\Omega_{\kappa}}
\newcommand{\omch}{\Omega_c h^2}
\newcommand{\ombh}{\Omega_b h^2}
\newcommand{\sig}{\sigma_8}

\begin{document}

\title{Constraining Perturbative Early Dark Energy with Current Observations}

\author{Ujjaini Alam}
\affil{ISR-1, ISR Division, Los Alamos National Laboratory, Los Alamos, NM 87545, USA}

%\thispagestyle{empty}

%\sloppy

\begin{abstract}
\small{ 
In this work, we study a class of early \de~(EDE) models, in which,
unlike in standard \de~models, a substantial amount of \de~exists in
the matter-dominated era. We self-consistently include \de~\perts, and
constrain these models using current observations. We consider EDE
models in which the \de~equation of state is at least $\wm \ggeq -0.1$
at early times, which could lead to a early \de~density of up to
$\Omega_{DE} (z_{CMB})= 0.03 \om (z_{CMB})$. Our analysis shows that,
marginalizing over the non-\de~parameters such as $\om, H_0, n_s$,
current CMB observations alone can constrain the scale factor of
transition from early \de~to late time \de~to $\at \ggeq 0.44$ and
width of transition to $\dt \lleq 0.37$. The equation of state at
present is somewhat weakly constrained to $\w \lleq -0.6$, if we allow
$H_0 < 60$ km/s/Mpc. Taken together with other observations, such as
supernovae, HST, and SDSS LRGs, $\w$ is constrained much more tightly
to $\w \lleq -0.9$, while redshift of transition and width of
transition are also tightly constrained to $\at \lleq 0.19, \dt \lleq
0.21$. The evolution of the equation of state for EDE models is thus
tightly constrained to $\ld$CDM-like behaviour at low
redshifts. Incorrectly assuming \de~\perts~to be negligible leads to
different constraints on the equation of state parameters-- $\w \lleq
-0.8, \at \lleq 0.33, \dt \lleq 0.31$, thus highlighting the necessity
of self-consistently including \de~\perts~in the analysis. If we allow
the spatial curvature to be a free parameter, then the constraints are
relaxed to $\w \lleq -0.77, \at \lleq 0.35, \dt \lleq 0.35$ with
$-0.014 < \omk < 0.031$ for CMB+other observations. For perturbed EDE
models, the $2\sigma$ lower limit on $\sig$ ($\sig \geq 0.59$) is much
lower than that in $\ld$CDM ($\sig \geq 0.72$), thus raising the
interesting possibility of discriminating EDE from $\ld$CDM using
future observations such as halo mass functions or the
Sunyaev-Zeldovich power spectrum. }
\end{abstract}

\maketitle

\section{Introduction}\label{intro}

Over the last decade, the unexpected faintness of distant Type Ia
supernovae have shown that the expansion of the universe is
accelerating at present
\citep{accl1, accl2, accl3, accl4, accl5, accl6, accl7, accl8, const, sdss_sne}. This remarkable discovery points to the existence of
\de~(DE), a negative pressure energy component which dominates the
energy content of the universe at present. Other, complementary,
probes such as the Cosmic Microwave background (CMB) and various large
scale structure surveys have also confirmed the existence of this
mysterious component of energy \cite{bao, wmap5, lrg}. Several
theories have been propounded to explain this phenomenon, the simplest
of which is the cosmological constant $\ld$, with a constant energy
density and a constant equation of state $w=-1$.  The cosmological
constant is fit well by the current data \citep{const}, however, there
are no strong constraints on the time evolution of \de~at
present. Thus, evolving models of \de~remain viable as alternative
candidates for dark energy. Many non-cosmological constant
phenomenological explanations for cosmic acceleration have been
suggested \citep[see reviews][and references
therein]{rev1, rev2, rev3, rev4, rev5, rev6, rev7}. These are based either on the
introduction of new physical fields (quintessence models, Chaplygin
gas, etc.), or on modifying the laws of gravity and therefore the
geometry of the universe (scalar-tensor gravity, $f(R)$ gravity,
higher dimensional `Braneworld' models \etc). As of now, there is no
consensus on the true nature of \de.

An interesting class of models which have been suggested in the
literature are early \de~models, a class of \de~in which the early
universe contained a substantial amount of \de. These models were
studied theoretically in \citep{ede1,ede2,ede3} and references
therein, and have been analyzed with respect to observations
extensively in recent times in \citep{ede4,ede5,ede6,ede7,ede8}. For
now, there are no strong observational constraints on the EDE models,
and it is especially difficult to discriminate EDE models which have
$w = -1$ at present from the $\ld$CDM model of \de.

In this work we use a parameterization of the equation of state of
\de~to study and constrain EDE models using the currently available
data. We attempt to see if bounds can be put on the transition from
early \de~to the present day \de~content of the
universe. Section~\ref{meth} explains the methods and data used for
this analysis, section~\ref{res} shows the results, and in
section~\ref{concl} we conclude.

\section{Methodology}\label{meth}

\De~\perts~for dynamic \de~models have been studied in a number of
works, usually under the formalism of a minimally coupled scalar field
\citep[See][and references therein]{pert1, pert2, pert3, pert4, pert5,
  pert6, pert7, pert8, pert9, pert10}.  For practical purposes, \eg analyzing dynamic
\de~models in light of data, it has sometimes been the practice to
consider \de~\perts~as negligible, and sometimes not. As shown in
\citep{quint_np}, not taking into account the \de~\perts~correctly can
lead to erroneous. gauge-dependent results. In our analysis, we
self-consistently include the \de~\perts~for the EDE models, and also
show how the results differ if these \perts~are not included.

\subsection{\DE~\Perts}\label{pert}

A homogeneous and isotropic large scale universe can be described by
the Friedman-Lemaitre-Robertson-Walker (FLRW) metric
\beq
ds^2 = a^2(\eta) \left[ d\eta^2 + \delta_{\alpha\beta}dx^{\alpha} dx^{\beta} \right] \,\,,
\eeq
where $\eta$ is the conformal time, $d{\bf x}$ is the length element,
and $a(\eta)$ is the scale factor. The speed of light $c$ is set to
unity, so that the time variable has dimensions of length.

First order \perts~take the form
\beq
ds^2 = a^2(\eta) \left[(1+2\Psi({\bf x}, \eta)) d\eta^2 - (1+2\Phi({\bf x}, \eta)) \delta_{\alpha\beta}dx^{\alpha} dx^{\beta} \right]\,\,,
\eeq
where $\Phi, \Psi$ are the Bardeen potentials. If proper isotropy of
the medium is zero, then $\Phi = -\Psi$.

We adopt two equivalent approaches to account for the \de~\perts, the
first consists of considering the \de~component as an additional
fluid, while in the second \de~is defined as a minimally coupled
scalar field. Both approaches lead to the same result within the
framework considered, and each has its usefulness in analyzing the
results.

\subsubsection{\DE~as a Fluid}\label{fluid}

In this section we follow the treatment of \cite{pert5}.  Along with
the matter and radiation components, we consider dark energy to be an
additional fluid component, so that the dark energy \perts~are
characterized by an equation of state, an adiabatic sound speed and an
intrinsic entropy \pert--
\ber
\ww &=& \frac{p_{DE}}{\rho_{DE}} \\
\ca &=& \frac{\dot{p}_{DE}}{\dot{\rho}_{DE}} \\
\Gamma_{DE} &=& \frac{\delta p_{DE}}{p_{DE}} -\frac{\ca}{\ww} \delta_{DE} \,\,.
\eer
Defining the frame invariant quantity $c_{s,i}^2$ (the fluid sound
speed in the frame comoving with the fluid), the continuity and Euler
equations giving the evolution of the density contrast and velocity of
a fluid with equation of state $w_i = p_i/\rho_i$, and adiabatic speed
of sound $c_{a,i}^2 = \dot{p}_i/\dot{\rho}_i$, may be written as
(prime denotes derivative with respect to $\eta$)
\ber\label{eq:pert}
\delta_i^{\pr} &=& -3 \HH (c_{s,i}^2-w_i) \delta_i - 9 \HH^2 (c_{s,i}^2-c_{a,i}^2)  (1+ w_i) \frac{v_i}{k}  \nonumber\\
&& -(1+ w_i) k v_i - 3(1+ w_i) \Psi^{\pr} \\
v_i^{\pr} &=& - \HH (1-3 c_{s,i}^2) v_i + \frac{k c_{s,i}^2 \delta_i }{(1 + w_i)} - kA  \label{eq:pert2}\,\,, 
\eer
where $A$ is the acceleration ($A=0$ in the synchronous gauge, $A=
-\Psi$ in the Newtonian gauge), and $\HH = a^{\pr}/a = a H$ is the
conformal Hubble parameter. For the matter component, $w_m = c_a^2 =
c_s^2 = 0$. For the dark energy component, a fluid with varying $\ww
\geq -1$ has $\ca = \ww - [d\ww/d({\rm ln} \ a)]/3(1+\ww)$. For scalar
field like \de~models, $\cs = 1$. For a more general class of models,
such as k-essence, $\cs$ could be variable as well. To reduce the
number of parameters, we consider $\cs = 1$ in our analysis, which
would still allow us to study a wide range of \de~models. Thus for an
universe containing matter (CDM+baryons) and dark energy, a set of
four \pert~equations may be defined for the gauge-independent
variables $\delta_m, v_m, \dlt, \vel$ and solved using adiabatic
initial conditions.

\subsubsection{\DE~\Perts~in Scalar Field Formalism}\label{quint}

An equivalent and convenient approach for studying the \de~\perts~is
to regard the dark energy component as a a minimally coupled scalar
field $Q$ with self-interaction potential $V(Q)$. The field dynamics
are given by
\beq
Q^{\pr\pr} + 2 \HH Q^{\pr} +a^2 \frac{d^2V}{dQ^2} = 0 \,\,,
\eeq
and the \perts~of the scalar field evolve through the perturbed
Klein-Gordon equation
\beq\label{eq:2q}
\delta Q^{\pr \pr} +2 \HH \delta Q^{\pr} + \left( k^2 + a^2 \frac{d^2 V}{dQ^2} \right) \delta Q = 4 Q^{\pr} \Psi^{\pr} - 2 a^2 \frac{dV}{dQ} \Psi \,\,.
\eeq
The metric \perts~evolve as
\beq\label{eq:2psi}
\Psi^{\pr \pr} + 3 \HH \Psi^{\pr} + 8 \pi G  a^2 V \Psi = 4 \pi G \left(Q^{\pr}\delta Q^{\pr} -a^2 \frac{dV}{dQ} \delta Q \right) \,.
\eeq
The matter density contrast may be obtained from the above
equations to be--
\ber\label{eq:2dm}
\delta_m &=& - \frac{1}{4 \pi G \rho_m} \left[ 3 \HH \Psi^{\pr} + \left \lbrace 8 \pi G a^2 (\rho_m + V) + k^2 \right \rbrace  \Psi \right. \nonumber\\
&& \left. + 4\pi G \left (Q^{\pr}\delta Q^{\pr} + a^2 \frac{dV}{dQ} \delta Q \right) \right ] \,\,.
\eer
The fluid parameters for the dark energy component (as defined in
section~\ref{fluid}) are related to the scalar field variables by
\ber
\ww &=& \frac{Q^{\pr 2} - 2 a^2 V(Q)}{Q^{\pr 2} + 2 a^2 V(Q)} \\
\ca &=& 1 + \frac{2}{3}\frac{dV}{dQ}\frac{a^2}{Q^{\pr} \HH} \\
\Gamma_{DE} &=& \frac{1-\ca}{\ww} [\dlt - 3\HH (1+\ww) \vel]\,\,,
\eer
and the gauge-independent \pert~variables by
\ber
\dlt &=& \frac{1}{a^2 \rho_{DE}} [Q^{\pr} \delta Q^{\pr} +a^2 \frac{dV}{dQ} \delta Q - Q^{\pr 2} \Psi ] \\
\vel &=& \frac{k \delta Q}{Q^{\pr}} \,\,.
\eer

\subsubsection{Imprint of \DE~on Observables}\label{de_obs}

The two basic \de~dependent observables are distance and growth
rate. Distance measures are based on standard candles, rulers, or
number densities as a function of redshift; growth rate measures are
based on density \perts~in linear theory. All distance measures are
ultimately based on the comoving distance to redshift $z$
\beq
r = \int_0^z \frac{dz}{H(z)} = \int_a^1 \frac{da}{a \HH(a)} \,\,, 
\eeq
\eg the SNe Type I a observations measure the magnitude of distant
SNe, given by $m_B(z) = 5 {\rm log}_{10} [(1+z) r(z)] + {\cal M}$.
The effect of \de~for distance measures is through the background
expansion of the universe, \ie, from the Hubble parameter $H(z) =
\HH(a)/a$. For CMB data, this comes in through the angular diameter
distance and the sound horizon
\ber
D_A(a) &=& a \int_1^a \frac{da}{a^2 H(a)} = a \int_1^a \frac{da}{a \HH (a)} \\
s(a) &=& \int_0^a \frac{c_s(a) \ da}{a^2 H(a)} = \int_0^a \frac{c_s(a) \ da}{a \HH (a)} \,\,.
\eer

The density \perts~are affected by the presence of \de~firstly through
the Hubble parameter, and secondly through the linear \pert~of \de, as
in eqs~(\ref{eq:pert}, \ref{eq:pert2}), or eqs~(\ref{eq:2q},
\ref{eq:2psi}, \ref{eq:2dm}). For the CMB power spectrum, the effect
of these is felt most strongly in the ISW effect at low $l$, as well
as in a shift of the peak positions.

The low $l$ observations can be understood as follows.  The behaviour
of the temperature anisotropy power spectrum in the CMB is given by
the covariance of the temperature fluctuation expanded in spherical
harmonics
\beq\label{eq:potl}
C_l = 4 \pi \int \frac{dk}{k} {\cal P}_x  |\Delta_l(k, \eta_0)|^2 \,\,,
\eeq
where ${\cal P}_x$ is the initial power spectrum, $\eta_0$ is the
conformal time today, and $\Delta_l(k, \eta_0)$ is the transfer
function at each $l$.

On large scales the transfer functions are of the form
\beq\label{eq:dpotl}
\Delta_l(k, \eta_0) = \Delta_l^{\rm LSS}(k) + \Delta_l^{\rm ISW}(k) \,\,,
\eeq
where $\Delta_l^{\rm LSS}(k)$ are the contributions from the last
scattering surface from the ordinary Sachs-Wolfe effect and
temperature anisotropy, and $\Delta_l^{\rm ISW}(k)$ is the
contribution due to the change in the potential $\phi$ along the line
of sight and is called the integrated Sachs-Wolfe (ISW) effect. The
ISW contribution can be written as
\beq\label{eq:isw}
\Delta_l^{I\rm SW}(k) = 2\int d\eta \ e^{-\tau(\eta)} \phi^{\prime} j_l[k(\eta-\eta_0)] \,\,,
\eeq
where $\tau(\eta)$ is the optical depth due to scattering of the
photons along the line of sight, and $j_l(x)$ are the spherical Bessel
functions.

The frame-invariant potential $\phi$, defined in terms of the Weyl
tensor, is equivalent to the Bardeen potential in the absence of
anisotropic stress and given by the Poisson equation
\beq
k^2 \phi = -4 \pi G a^2 \overline {\delta \rho} \,\,,
\eeq
while its derivative in a matter plus \de~universe, which is the
source term for the ISW contribution, is given by
\beq
k^2 \phi^{\pr} = -4 \pi G \frac{\partial}{\partial \eta} \left[ a^2 (\overline{\delta\rho_m} + \overline{\delta\rho_{DE}}) \right ] \,\,.
\eeq
From the above equations, it is clear that the magnitude of the ISW
contribution is dependent on the late time evolution of the total
density \perts, therefore on the \de~\perts.  It should be noted
however, that these are not independent of other cosmological
parameters, and the effect of \de~could be masked due to the
degeneracy of the \de~parameters with other parameters such as $H_0$
and the curvature of the universe.

\subsection{Parameterization of Equation of State of \DE}\label{parm}

To study EDE models under this formalism, we consider a
$w$-parameterization which may represent a large class of varying
\de~models \cite{coras} 
\beq\label{eq:wparm} 
w(a) =
\w+(\wm-\w)\frac{1+e^{\at/\dt}}{1+e^{(a-\at)/\dt}}\frac{1-e^{(a-1)/\dt}}{1-e^{1/\dt}}\,\,, 
\eeq 
where $\w$ is the equation of state of dark energy today, $\wm$ is the
equation of state in the matter dominated era, $\at$ is the scale
factor at which the transition between $\w$ and $\wm$ takes place, and
$\dt$ is the width of the transition.  If $\wm$ is allowed to be a
free parameter this parameterization can encompass a large class of
models, including $\ld$CDM and $w =$ constant models.  Models with
constant or slowly varying $w \simeq -1$ would be consistent with
current observations, however these are not EDE models, as they have
negligible amounts of \de~at early times.  For such models, there
would be very poor constraints on the transition parameters, since no
significant transition takes place between early time and late time
\de. Allowing these models in the analysis would therefore cause the
constraints on $\at, \dt$ to weaken. Leaving the amount of early
\de~free would be interesting when comparing EDE models with $\ld$CDM
and other \de~models. Such comparisons have previously shown that
while it is possible to put an upper limit on the amount of early \de,
it is not possible to put strong constraints on the evolution of
\de~if all the different \de~models are considered.  Previous studies
\citep{ede3, ede8} have constrained early time dark energy density to
$\simeq 3 \%$ of the matter density, however, as seen in \cite{ede8},
the evolution of \de~is weakly constrained. In this work, we study the
EDE models exclusively, to put constraints on the transition from
early to late time \de. If we are able to constrain the minimum
redshift (or maximum scale factor) at which such a transition occurs,
we would know that any signature for EDE would be found only in
observations beyond that redshift. This would also put a constraint on
the evolution of \de~at low redshifts. For studying EDE models with
this parameterization, we therefore choose $\wm > -0.1$, to ensure the
presence of adequate amounts of \de~at early times, so that we may put
constraints on the transition from early to late-time \de~for these
models.

\subsection{Observations}\label{obs}

We use the latest version of COSMOMC \cite{cosmomc} for our analysis,
modifying the CAMB module, as well as the various modules pertaining
to large scale structure and supernova observations in COSMOMC, using
the equations defined in section~\ref{pert}.  For the analysis using
only CMB data, we use the 5 yr WMAP \citep{wmap5}, CBI \citep{cbi},
VSA \citep{vsa}, BOOMERANG \citep{boom} and ACBAR \citep{acbar}
datasets.  In addition to the CMB data, we use other observations as
well. For supernovae, we use the Constitution dataset (SALT)
\citep{const}. This dataset comprises of 397 Type Ia SNe, of which
about 200 are at redshifts $z \lleq 0.1$, and the remaining are
distributed between $z = 0.1$ and $z = 1.7$. We also use the latest
SDSS data release (DR7) luminous red galaxy (LRG) data \cite{lrg}, and
the recent value of the Hubble constant from the SHOES (Supernovae and
$H_0$ for the Equation of State) program, $H_0 = 74.2 \pm 3.6$
km/s/Mpc ($1\sigma$) \citep{shoes}, which updates the value obtained
from the Hubble Key Project \citep{hst}. We incorporate a top-hat
prior on the age of the Universe, $10 \ {\rm Gyr} < t_0 < 20 \ {\rm
  Gyr}$. The addition of these other observations allows us to
constrain parameters such as $H_0$ which might otherwise be degenerate
with the \de~parameters of interest to us.

\section{Results}\label{res}

We first study the effect of the different \de~components on the
observations. To this purpose we choose two \de~models-- (i) a
\de~model with constant equation of state $\ww = -0.9$ (ii) an EDE
model with $\w = -1.0, \wm = -0.1, \at = 0.3, \dt = 0.2$. We compare
the behaviour of these two models of \de~with that of a $\ld$CDM
($\ww=-1$) model. All three models have identical values for the
non-\de~cosmological parameters (\eg $\omt, H_0$). The first model is
chosen for comparing the behaviour of non-perturbative and
perturbative \de~for a \de~model close to the cosmological constant in
behaviour, while the second is chosen for specifically studying how
early \de~affects the results.

\subsection{Effect of \DE~\Perts~on Observable Quantities}\label{example}

\begin{figure*} 
\centering
\begin{center}
\epsfxsize=7.0in
\epsffile{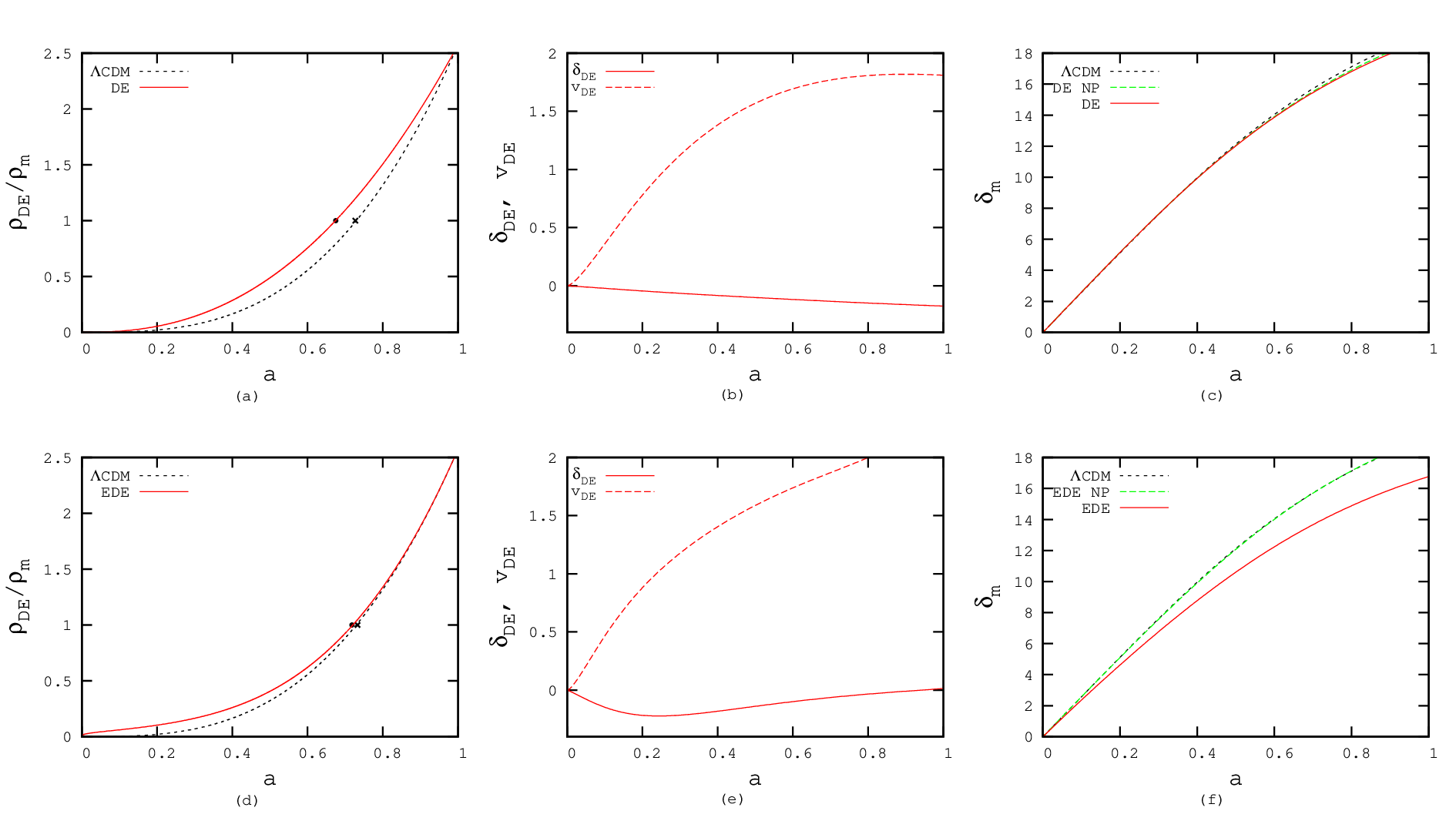}
\end{center}
\vspace{0.0cm}
\caption{\small 
Growth of relative \de~density $\rho_{DE}/\rho_m$ (panels (a), (d)),
\de~density contrast $\dlt$ \& velocity perturbation $\vel$ (panels
(b), (e)), and cold dark matter density contrast $\delta_m$ (panels
(c), (f)) with the scale factor $a$ for two \de~models : DE with $\ww
= -0.9$, and EDE with $\w = -1.0, \wm = -0.1, \at = 0.3, \dt = 0.2$
respectively. The red line in panels (a) and (d) represents the DE and
EDE models respectively, the black line in panels (a), (c), (d), (f)
shows the $\ld$CDM model for comparison, the \de~model without
perturbation is shown in green in panels (c) and (f), while the case
with perturbation is shown in red in these panels. The solid line in
panels (b) and (e) represents the \de~density contrast, while the
dashed line shows the velocity perturbation.  The filled circle in
panels (a), (d) represent matter-\de~equality for the DE and EDE
models respectively, while the cross represents matter-\de~equality
for $\ld$CDM.}
\label{fig:example}
\end{figure*}

Following \cite{pert5} we look at the effect on observations using the
\de~as a fluid framework. We first study the effect of
non-perturbative \de~on the observations. In an universe containing
matter and a smooth \de~component, the matter \perts~may be calculated
from Eqs~(\ref{eq:2q}, \ref{eq:2psi}, \ref{eq:2dm}), using $\delta Q =
\delta Q^{\pr} = 0$, to be--
\beq\label{eq:nopert}
\delta_m^{\pr \pr} - \HH \delta_m^{\pr} -4 \pi G \rho_m \delta_m = 0 \,\,.
\eeq
From this equation, we see that the \de~component appears only in the
second term which is effectively a damping term, therefore a
non-negligible amount of smooth \de~would suppress the clustering of
matter at large scales. Thus the only effect of \de~for a smooth
\de~model arises through the \de~density, for both geometric (\eg Type
Ia SNe) and perturbative (\eg CMB, matter power spectrum) data. For
matter-dominated regime, the above equation would result in $\delta_m
\propto a$. For the DE model with a constant $\ww > -1$, the
transition between matter and \de~happens earlier than for $\ww = -1$,
and more slowly, thus constant $\ww > -1$ models are expected to have
a smaller contribution to the ISW effect than $\ld$CDM. In
fig~\ref{fig:example} (a), we show the expansion history of the DE
model considered as well as that of $\ld$CDM. We see that the
\de~density equals matter density earlier in the \de~model, and we
expect this to have a noticeable effect in the scalar $C_l$'s for CMB
data. For the EDE model, things are slightly different, as seen in
figure~\ref{fig:example} (d). Since the value of the equation of state
today is $\w = -1$, the transition from matter to dark energy occurs
at nearly the same time as on $\ld$CDM. Also, because initially the
\de~density is higher in this model, this transition is
flatter. Therefore we may expect that these models would have a larger
contribution to the ISW effect. The effect on the matter \perts~is a
mild suppression for both cases as expected from eq~(\ref{eq:nopert}),
seen in figure~\ref{fig:example} (c), (f).

Setting the \de~\perts~to zero artificially is however not consistent
with the general relativity framework except in the case of a
cosmological constant, $\ww = -1$. We therefore now add the
\de~\perts~to the calculation. We consider the gauge comoving with
dark matter, in which the acceleration is zero. If $\dlt$ is initially
zero, we see from eq.~(\ref{eq:pert}) that it is sourced by the other
perturbations if $\ww \neq -1$ via the the source term $3(1+\ww)
\Psi^{\pr}$. An over density causes a decrease in the local expansion
rate so that $\Psi^{\pr} < 0$. In this case a fluid starts to fall
into overdensities if $\ww > -1$. In the subsequent evolution of
\de~\perts, if $\cs =1$, then the source term for the velocity $\vel$
is positive, thus causing the velocities to be anti-damped.  For the
density contrast $\dlt$, when $k << \HH$, the term $(1+\ww) k \vel$
can be neglected and the velocity and wavenumber enter only via the
combination $(1+\ww)\vel/k$, which is small. Thus the evolution of
$\dlt$ is almost $k$-independent at large scales, and the two
remaining source terms $-3 \HH (\cs - \ww) \dlt$ and $- 3
(1+\ww)\Psi^{\pr}$ are of opposite signs with $\dlt > 0$ and
$\Psi^{\pr} < 0$ initially. Therefore \de~\perts~change sign at very
early times and start decreasing, having the opposite sign to that of
the matter \perts, which source $\dlt$ through the now increasing
$\Psi^{\pr}$. For the DE model with constant $\ww$, we see this effect
in figure~\ref{fig:example} (b). Thus $\dlt$ and $\delta_m$ have
opposite signs, and at late times when the \de~becomes a significant
fraction of the energy density, the total density \perts~are
\emph{smaller} than those without \de~\perts. So there is a larger
overall change in the potential $\phi^{\pr}$ in eq~(\ref{eq:dpotl}),
and the ISW contribution is increased. Since the total decrease in
$\dlt$ is small as $\ww$ is close to the $\ld$CDM value of $-1$, the
matter \perts~do not change significantly, as seen in
fig~\ref{fig:example} (c). For a DE model with $\ww >> -1$, the effect
on $\delta_m$ would be stronger.

For EDE, varying $\ww$ provides a further effect. Initially when $\wm
= -0.1$, the source terms approximate to $-3 \Psi^{\pr}$ and $-3 \HH
\dlt$, and since $\Psi^{\pr}$ is significantly larger than $\dlt$, it
is the primary source term in eq~(\ref{eq:pert}). Therefore, $\dlt$
decrease rapidly, more than it would for $\ww \sim -1$. When the
\de~equation of state transitions from $\wm = -0.1$ to $\w = -1.0$ at
$a_t = 0.2$, the source term $-3 \HH (\cs - \ww) \dlt$ becomes larger
and therefore the decreasing $\dlt$ starts to increase, though not
fast enough to change signs again, as seen in figure~\ref{fig:example}
(e). $\dlt$ is therefore still of opposite sign to $\delta_m$, but
less negative than for a $\ww = {\rm constant} \sim -1$ case. Thus the
ISW contribution is decreased from what it would be in the no
perturbation case, but still is larger than that for the $\ld$CDM
model, while the matter perturbations at low $k$, which source the
\de~\perts~through $3(1+\ww) \Psi^{\pr}$, become smaller at late times
as $\ww$ becomes more negative. Thus matter perturbations at low $k$
for EDE models are strongly suppressed at late times as compared to
$\ld$CDM, or the no perturbation case (fig~\ref{fig:example} (f)). The
change in potential $\phi^{\pr}$ in eq~(\ref{eq:dpotl}) is therefore
enhanced. So effectively, we expect a strong enhancement off the
transfer function and therefore the matter power spectrum at large
scales (low $k$). Thus the matter power spectrum at late times, when
normalized at low $k$, would show a strong suppression on the small
scales (\ie~at high $k$), and this suppression is effected due to the
variation of the \de~equation of state.

The effect of \de~\perts~can be understood also from the scalar field
formalism. From eq~(\ref{eq:2q}), the scalar field $Q$ can be viewed
as a fluid with comoving Jeans mode given by the curvature of the
potential, i.e. the mass of the field, $k_J = a
\sqrt{d^2V/dQ^2}$. Therefore scales which corresponds to modes $k <
k_J$ will collapse under gravitational instability, while modes $k >
k_J$ will undergo a series of damped oscillations due to pressure
waves in the quintessence fluid. This has two major effects. Firstly,
the large scale clustering of \de~enhances the amplitude of the ISW
effect in CMB at low $l$. Secondly, as a consequence of the
homogeneity of of the \de~component on small scales and the fact that
the growth of the linear matter perturbations is suppressed due to the
lower values of $\omt$, the linear matter power spectrum at small
scales will have an amplitude which is smaller than in $\ld$CDM. We
thus expect that on the very large scales ($k < k_J$ ) the dark energy
clustering enhances the matter power spectrum compared to the
unclustered case, while on small scales ($k > k_J$ ) the opposite
occurs. If we CMB normalize the matter power spectrum (\ie~normalize
it at large scales), the small scale matter power spectrum will show a
stronger suppression of power than in the no \pert~case, thus giving a
smaller value of $\sig$ at present.

\begin{figure*} 
\centering
\begin{center}
\epsfxsize=7.1in
\epsffile{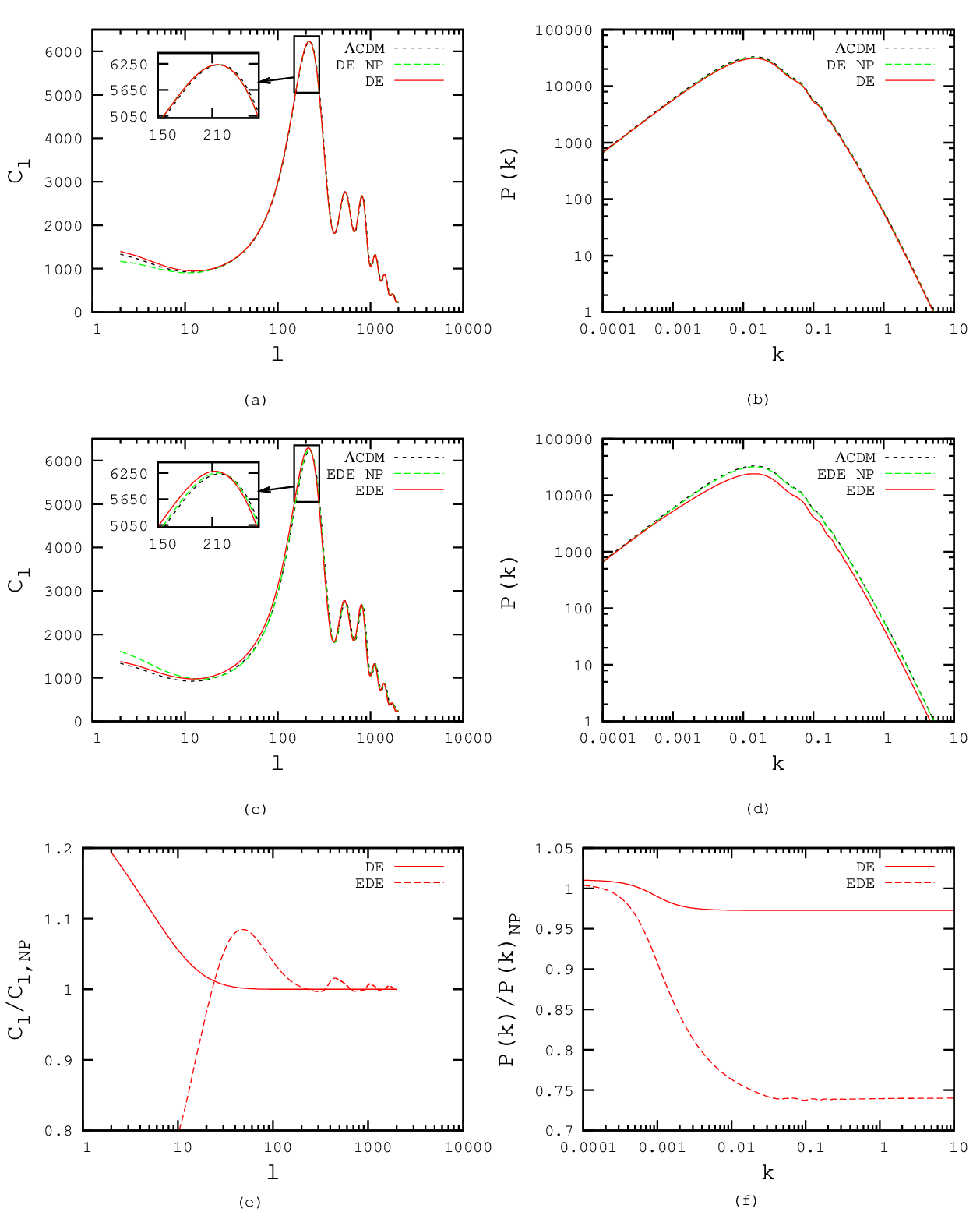}
\end{center}
\vspace{0.0cm}
\caption{\small 
Scalar $C_l$'s and matter power spectrum for the DE model with $\ww =
-0.9$ (panels (a) and (b)), and the EDE model with $\w = -1.0, \wm =
-0.1, \at =0.3, \dt -=0.2$ (panels (c) and (d)). The black line in each
panel represents the corresponding $\ld$CDM model, the green line
represents the \de~model with no \de~\perts, while the red line
represents the case with \de~\perts~taken into account. The insets in
the panels (a) and (c) show the shift in the position of the first
peak for the \de~model considered. Panels (e) and (f) show the
difference between the perturbed and unperturbed cases for both the DE
(solid line) and the EDE (dashed line) models for the two observables.
}
\label{fig:obs}
\end{figure*}

Fig~\ref{fig:obs} (a), (b) show the CMB $C_l$'s and the matter power
spectrum at $z = 0$ normalized to CMB for the DE model. As expected
from the arguments in the previous paragraphs, we see that there is a
slight shift in the CMB peak position as well as enhanced power at low
$l$ for the DE model as compared to $\ld$CDM. The main effect is at
low $l$, a region which is cosmic variance limited, therefore
difficult to rule out observationally. For the matter power spectrum,
as expected, there is a small suppression of power at high $k$ (since
the normalization is done at low $k$). The value of $\sig$ in the no
\pert~case is $\sig = 0.79$, while that in the perturbed case is $\sig
= 0.80$, and that for $\ld$CDM is $\sig = 0.82$. Neither the effect on
CMB nor that on the matter power spectrum is in itself good enough to
rule out the DE model, even for the case where DE perturbations have
been accounted for.  For the EDE model, as seen in fig~\ref{fig:obs}
(c) (d), the non-perturbative case shows effect mostly in the low $l$
regime through the ISW effect, which is cosmic variance limited. The
results for the matter power spectrum today also show a very slight
difference from the cosmological constant. These results appear to
suggest that just the non-perturbative effects of \de~are not
sufficient to discriminate this EDE model from $\ld$CDM, especially if
we factor in degeneracies with other cosmological parameters, such as
$H_0$. When we consider the perturbative case, the ISW effect is
actually muted, however, there is a slightly larger shift in the CMB
peak position, (see inset of fig~\ref{fig:example} (c)) which is a
tightly constrained observable. The matter power spectrum at present
shows a stronger suppression at small scales which leads to a much
smaller value of $\sig = 0.69$ (as compared to the non-perturbative
case, where $\sig = 0.81$, which is close to the $\ld$CDM
value). Thus, although the background expansion of this model is very
similar to $\ld$CDM at late times, its early time behaviour leaves
signatures for discriminating it from the $\ld$CDM model provided the
\de~\perts~are accounted for properly. The effect of adding the
\de~\perts~is seen in fig~\ref{fig:obs} (e), (f) for both DE and EDE
models. In obtaining the scalar $C_l$s, for the DE model, there is a
fairly large difference at low $l$, while at high $l$ the perturbed
and non-perturbed models behave similarly. For the EDE model, there is
a large difference at low $l$, and also a significant difference at
the higher $l$s. For the matter power spectrum today, the EDE model
shows a larger difference in in the perturbed and non-perturbed
case. Thus, a model close to $\ld$CDM today as also in the past (as in
the DE model chosen) would be difficult to discriminate from $\ld$CDM
from current observations, but a model with a different expansion
history in the past, even if it is very similar to $\ld$CDM today
(such as the EDE model), could be discriminated using the perturbative
observations such as CMB and the matter power spectrum provided the
\de~\perts~are not neglected. These results are commensurate with
those found in \citep{pert1} where constant equation of state models
of \de~were considered, and those in \citep{pert11}, where quintessence
models of \de~were studied.

\begin{figure*} 
\centering
\begin{center}
\epsfxsize=3.6in
\epsffile{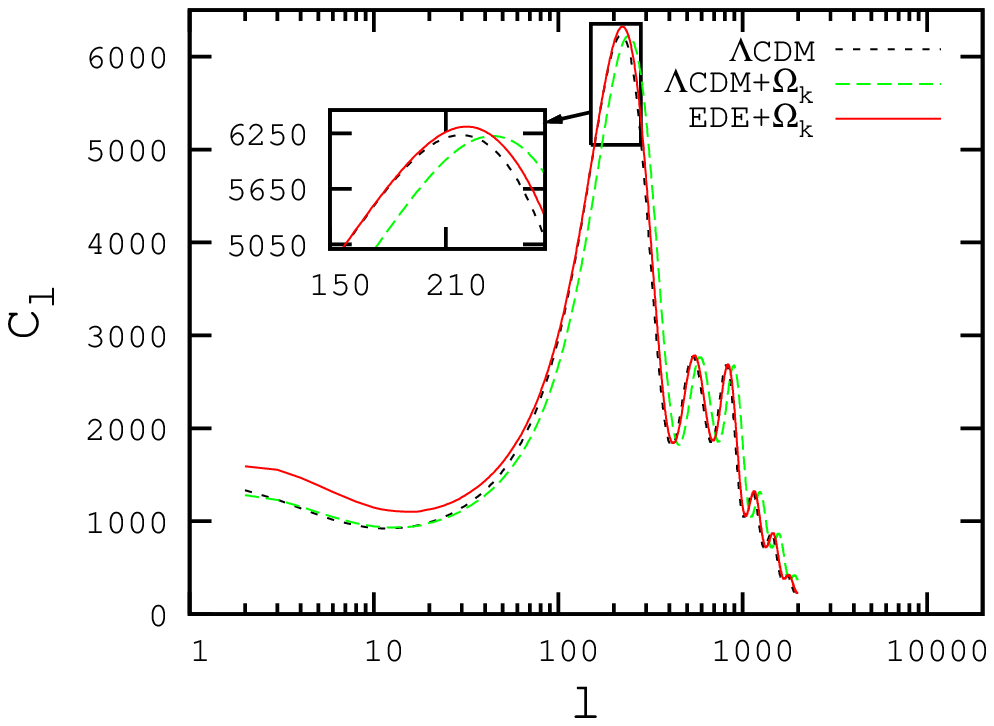}
\end{center}
\vspace{0.0cm}
\caption{\small 
Scalar $C_l$'s for $\ld$CDM and EDE models with curvature. The black
line in each panel represents the flat $\ld$CDM model, the green line
represents $\ld$CDM with $\omk = 0.06$, and the red line represents an
EDE model with $\w = -0.65, \wm = -0.1, \at = 0.2, \dt = 0.1$, with
$\omk = 0.06$. The inset shows the shift in the position of the first
peak for the $\ld$CDM model with curvature. For the EDE model with
curvature, the shift in the first peak is compensated, however, the
height of the first peak as well as the low-$l$ behaviour differs from
the flat $\ld$CDM case.}
\label{fig:curv}
\end{figure*}

We note here that, since in addition to the ISW effect, dark energy
also makes itself felt in a shift of the CMB first peak position, we
expect that the \de~parameters may be degenerate with $\omk h^2$ if
the flatness condition is removed in the analysis. We study the effect
of curvature on the scalar $C_l$'s in figure~\ref{fig:curv}. A
non-flat $\ld$CDM model will differ from a flat $\ld$CDM model with
all other parameters identical mainly in a shift of the peak
positions. Figure~\ref{fig:curv} shows this shift for a $\ld$CDM model
with $\omk = 0.06$. An EDE model with $\w = -0.65, \wm = -0.1, \at =
0.2, \dt = 0.1$, and a curvature $\omk = 0.06$ is also shown. For the
EDE model, the dark energy component compensates for the curvature of
the universe, thus the peak position is the same as for the flat
$\ld$CDM model. However, as seen in the previous paragraphs, EDE
manifests itself not only in the shift of the peaks, but also in the
shape of the peaks and in the low-$l$ ISW effect. In this example, the
height of the first peak is different for the EDE model, as is the low
$l$ behaviour, rendering it distinct from the flat $\ld$CDM
model. Thus, although we expect some degeneracy between the
\de~parameters and the curvature, this degeneracy is not very strong,
since both the position and the height of the first peak are strongly
constrained by current CMB data.

\subsection{Constraints from Observations}

We first study the results using only the CMB data. The primary
parameters to be varied are the standard CMB parameters-- $\omch,
\ombh, \theta, \tau, n_s, A_s$, and the equation of state parameters
$\w, \wm, \at, \dt$. Since we wish to study EDE models, we restrict
the equation of state at early times to $w_m \lleq 0.1$. This can give
rise to a \de~density of up to $\omde(z) \lleq 0.03 \om(z)$ at early
times.  We assume a flat universe, \ie~$\omk = 0$, and consider the
full \de~\perts. The secondary parameters that we deduce from the
analysis are $\omt, H_0, \omde/\om (z_{CMB}), \sig$. The first column
of table~\ref{tab:cmb} shows the mean and $2\sigma$ boundaries for the
primary and secondary parameters. We see that the EDE parameters are
constrained at $\w < -0.61, \at < 0.44$ (which means $z_t > 1.2$),
$\dt < 0.37$. The constraints on the scale and width of transition are
reasonable, however, the constraint on the equation of state today,
$\w$, is too broad. We note however, that this result is obtained by
using CMB data alone, using other data would reduce degeneracies with
the other parameters. For instance, SNe Type Ia data would affect the
equation of state today more strongly. Also, the Hubble parameter for
which $\w \simeq -0.6$ is allowed is $H_0 \simeq 60$ km/s/Mpc, much
lower than the currently accepted measurement for it
\cite{shoes}. Therefore, we expect that the addition of other
observations to the analysis should improve the constraints on the EDE
parameters significantly. It is interesting to note also that the
$2\sigma$ lower bound on $\sig$ for this analysis is as low as $\sig
\geq 0.49$, whereas the $\ld$CDM fit to the WMAP5 data has a $\sig
\geq 0.72$.

\begin{figure*} 
\centering
\begin{center}
\epsfxsize=6in
\epsffile{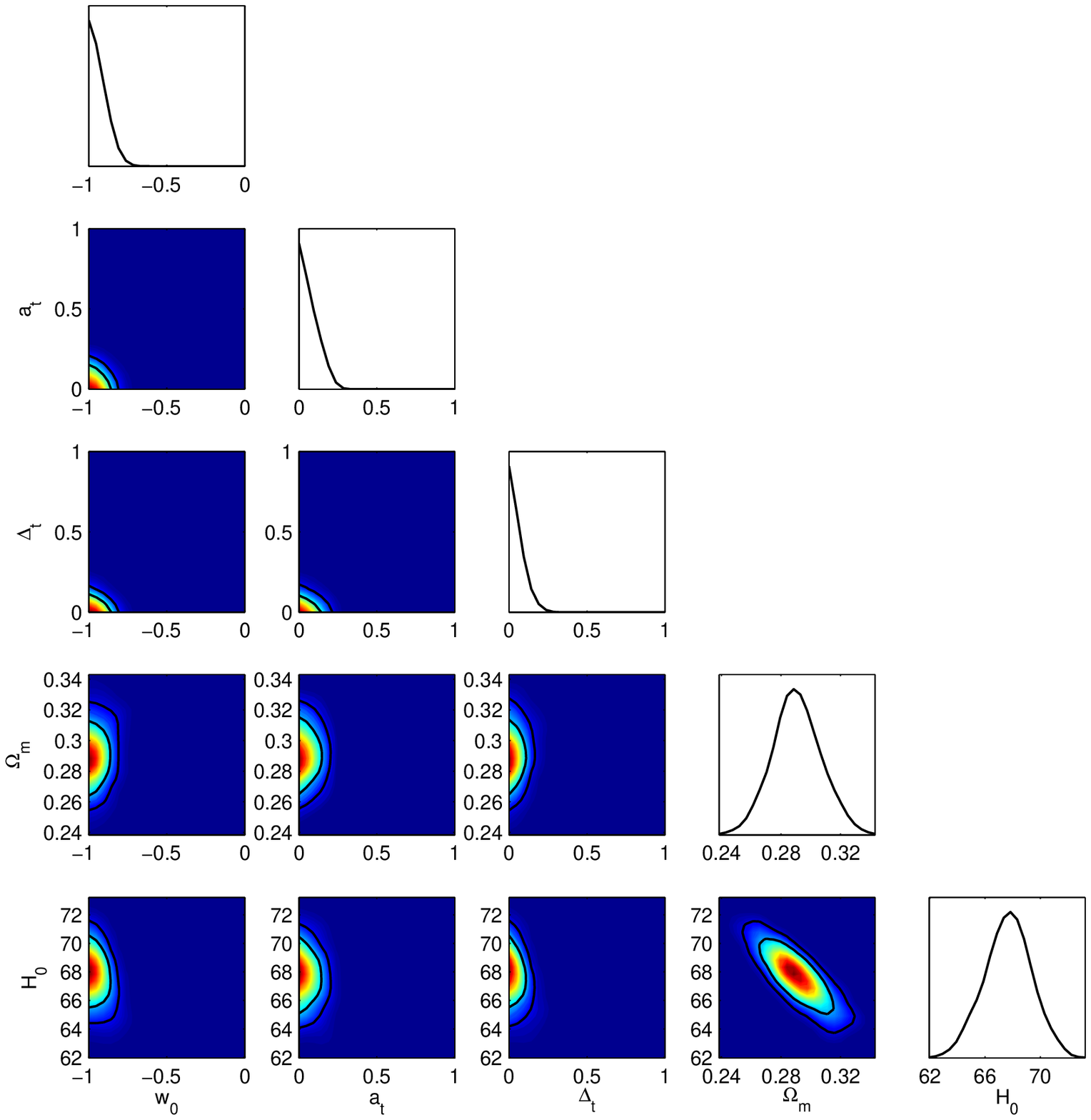}
\end{center}
\vspace{0.0cm}
\caption{\small 
Constraints from CMB (WMAP5, CBI, VSA, BOOMERANG, ACBAR) and other
datasets (SNe Type Ia Constitution, LRGDR7, SHOES) on EDE parameters
$\w, \at, \dt$, as well as $\omt, H_0$, showing marginalized
one-dimensional distributions and two-dimensional $68 \%$ and $95 \%$
limits. Full \de~\perts~ are taken into account and the curvature of
the universe is fixed at $\omk = 0$.}
\label{fig:alldat}
\end{figure*}

\begin{figure*} 
\centering
\begin{center}
\epsfxsize=3.6in
\epsffile{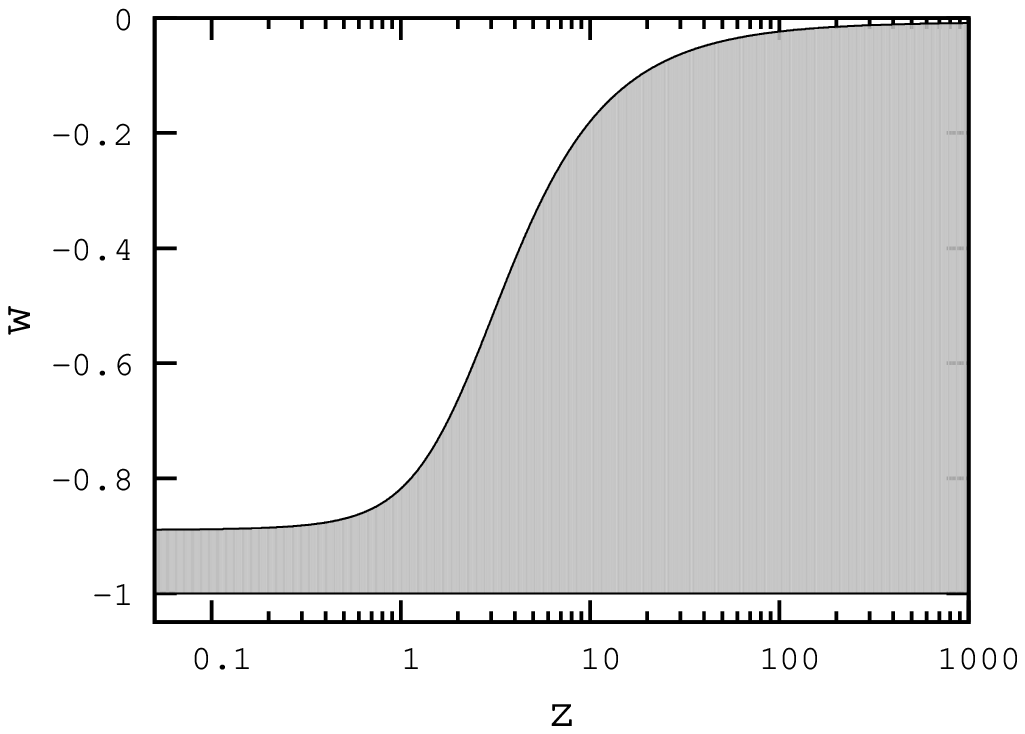}
\end{center}
\vspace{0.0cm}
\caption{\small 
$2\sigma$ confidence levels from CMB (WMAP5, CBI, VSA, BOOMERANG,
ACBAR) and other datasets (SNe Type Ia Constitution, LRGDR7, SHOES) on
the equation of state of \de, for the EDE models considered. Full
\de~\perts~ are taken into account and the curvature of the universe
is fixed at $\omk = 0$.}
\label{fig:w_cmb}
\end{figure*}

We now redo the analysis adding other datasets to see how the
constraints improve. Three distinct cases are considered-- (a) full
\de~\perts~are taken into account, $\omk = 0$; (b) \de~\perts~are
considered negligible, $\omk = 0$; and (c) full \de~\perts~are
considered, and the constraint on the flatness of the universe is
lifted (\ie~$\omk$ is a free parameter). The results are shown in the
second, third and fourth columns of table~\ref{tab:cmb}. For the fully
perturbed, flat case, when all the data is considered, the EDE
parameters are constrained to $\w < -0.89, \at < 0.19$ (\ie~$z_t >
4.2$), $\dt < 0.21$. The addition of other datasets clearly enhances
the constraints on the EDE model. This is because the other parameters
which could be degenerate with the EDE parameters, such as the Hubble
parameter, are well-constrained by other observations. We note that
the constraint on the equation of state today, $\w$, is stronger than
that would be obtained using the background data alone (\eg for Type
Ia SNe, we find $\w \lleq -0.75$ for constant equation of state, when
systematics are included \cite{const}). In figure~\ref{fig:alldat}, we
show the two-dimensional $68 \%$ and $95 \%$ confidence levels, as
well as the marginalized one-dimensional distributions for the EDE
parameters of interest, $\w, \at, \dt$, and the matter density $\omt$
and the Hubble parameter $H_0$, which are expected to be degenerate
with the EDE parameters. We see that all three EDE parameters are now
strongly constrained, and the non-EDE parameters are close to the
values expected in the $\ld$CDM model. The evolution of the equation
of state of \de~with redshift is shown in figure~\ref{fig:w_cmb}. We
see that at low redshifts ($z \lleq 2$), the $2\sigma$ confidence
level for $w(z)$ is quite close to $\ld$CDM. Thus current observations
already constrain the evolution of the equation of state for EDE
models to $\ld$CDM-like behaviour at present and in the near
past. Studying the background expansion data (which is usually below
redshift of two) will therefore not able to distinguish these EDE
models from $\ld$CDM with any success even if there is adequate
amounts of \de~at early times. In order to distinguish these EDE
models (currently accepted by the data) from $\ld$CDM, we need to look
at the perturbative data. Thus we may conclude that even if there is
significant amount of \de~in the universe at early times, this has to
reduce to \de~very close to $\ld$CDM at present times, that this
transition cannot take place too late (around redshift of four) and
that the transition needs to be sharp ($\dt \lleq 0.2$). However, we
should note that even with these constraints, $\sig$ is still
significantly different from the typical $\ld$CDM value, with the
$2\sigma$ lower bound being at $\sig \ggeq 0.6$. This is because, as
discussed earlier in section~\ref{example}, EDE has a strong effect on
the matter power spectrum, leading to a much lower $\sig$ than that in
the cosmological constant model. This means that studying data which
utilizes the matter power spectrum (such as the halo mass functions)
even at low redshifts may allow us to discriminate between EDE and
$\ld$CDM models.

We next look at the case where \de~\perts~are neglected, for the full
dataset. We find that, although the results are similar for many of
the parameters, they can be rather different for the EDE
parameters. As seen in the third column of table~\ref{tab:cmb}, the
EDE parameters are constrained to $\w < -0.8, \at < 0.33$ (\ie~$z_t >
2$), $\dt < 0.31$, thus, neglecting the perturbations for an EDE model
would result in rather broader constraints on its parameters. The
value of $\sig$ allowed at $2\sigma$ is also much closer to the
$\ld$CDM value, with $\sig > 0.72$. Neglecting the \de~\perts~in a
dynamic \de~scenario may therefore produce results very different from
the true results when full \de~\perts~are considered.

If the flatness of the universe constraint is removed, taking
\de~\perts~into account, the EDE parameters are mildly degenerate with the
curvature of the universe $\omk$. As seen in the last column of
table~\ref{tab:cmb}, the EDE parameters in this case are constrained
to $\w < -0.77, \at < 0.35$ (\ie~$z_t \ggeq 2$), $\dt < 0.35$, while
the curvature of the universe is still rather tightly constrained to
$-0.014 < \omk < 0.031$. Thus, relaxing the flatness constraint leads
to a weakening of the constraints on the parameters of the EDE models,
but can still lead to reasonable constraints on the EDE parameters.

Previous works that have studied EDE with perturbations have
constrained the amount of early \de~using current observations, \eg
\cite{ede8} obtained $\Omega_{EDE} < 1.4 \times 10^{-3}$. However, as
explained in section~\ref{parm}, this study allowed for models of
\de~that have negligible amounts of \de~at early times. This led to a
weakening of the constraints on the EDE transition parameters, with
the parameter $\wm$ attaining peaks both at $\simeq -1$ and $\simeq
0$, and the equation of state today being close to $\ld$CDM. Thus no
strong constraint could be put on the evolution of the equation if
state. In this work we have attempted to address the question of how
to put constraints on the transition of early time to late time \de~if
the universe contains a certain amount of early \de. If we constrain
early \de~to $\wm \ggeq -0.1$, we exclude models which do not have EDE
behaviour, and thus are able to put reasonable constraints on the
transition parameters, which give us an insight into the evolution of
\de~for these models. It is difficult to rule out the presence of EDE
altogether, due to the dearth of data at very high redshifts, but with
this study we are able to put constraints on when the universe could
have transited from such early time \de~to late time, $\ld$CDM-like
behaviour. We find that late time behaviour of the equation of state
of these models must be close to $\ld$CDM below redshift of
few. However, since the $\sig$ of these models is rather different
from $\ld$CDM, they may be distinguished from $\ld$CDM using data such
as the halo mass function, or the Sunyaev-Zeldovich power spectrum,
even at the lower redshifts. In addition, this work also studies the
degeneracy between the curvature of the universe and the
\de~parameters.

\section{Conclusions}\label{concl}

In this work, we have studied early \de~models using current
observations. We find that, if a sizeable amount of \de~exists in
early times ($\omde (z_{CMB}) \simeq 0.03 \om (z_{CMB})$), we may put
tight constraints on the transition of this \de~to its present day
value, and that the present day value of the \de~equation of state
must be close to the $\ld$CDM value. If the \de~\perts~are correctly
accounted for, then the current \de~equation of state is constrained
to $\w < -0.89$, while the transition from early \de~must occur at
redshifts of $z_t > 4.2$, with a narrow transition width of $\dt <
0.21$. Incorrectly assuming that \de~\perts~are negligible leads to a
different result-- $\w < -0.8, z_t > 2, \dt > 0.31$, thus showing that
it is vital to include the \de~\perts~self-consistently in any
analysis that uses perturbative data such as CMB or the matter power
spectrum. Leaving $\omk$ to be a free parameters leads to a weakening
of the constraints on the \de~parameters, with $\w < -0.77, \at <
0.35$ (\ie~$z_t \ggeq 2$), $\dt < 0.35$ for $-0.014 < \omk < 0.031
$. We note that, for the flat universe in which \de~\perts~are
considered, the value of $\sig$ is much lower than that in
corresponding $\ld$CDM models. As will be shown in a companion paper
\cite{ede_lss}, this may lead to interesting constraints from future
large scale structure data such as halo mass functions, as also from
the Sunyaev-Zeldovich power spectrum.

\section{Acknowledgements}

The author would like to thank S. Bhattacharya, J. Bullock, S. Habib,
K. Heitmann, M. Kaplinghat, Z. Lukic, A. Pope and S. Zoudaki for
interesting discussions, and the referee for useful comments. This
work was supported by the LDRD program at Los Alamos National
Laboratory.

\bibliographystyle{jphysicsB}

\newpage

\begin{table*}
\begin{center}
{\footnotesize 
\caption{\footnotesize 
Cosmological parameters and their $95 \%$ confidence level intervals
obtained from the analysis of CMB and other observations.}
\begin{tabular}{cccccccc}
\hline
&CMB only&&&CMB+other datasets&\\
\hline
&Pert;&&(a) Pert;&(b) No Pert;&(c) Pert;&\\
&$\omk = 0$&&$\omk = 0$&$\omk = 0$&$\omk$ free&\\
\hline
Primary Parameters&&&&&&\\
\\
$100 \ombh$&$2.271^{0.140}_{0.110}$&&$2.300^{0.128}_{0.170}$&$2.242^{0.131}_{0.116}$&$2.264^{0.131}_{0.120}$\\
\\
$\omch$&$0.110^{0.014}_{0.015}$&&$0.109^{0.012}_{0.010}$& $0.115^{0.007}_{0.013}$& $0.115^{0.008}_{0.014}$\\
\\
$\theta$&$1.042^{0.005}_{0.004}$&&$1.043^{0.004}_{0.007}$&$1.042^{0.005}_{0.006}$&$1.041^{0.007}_{0.006}$\\
\\
$\tau$&$0.079^{0.063}_{0.016}$&&$0.087^{0.045}_{0.040}$&$0.078^{0.037}_{0.023}$&$0.075^{0.026}_{0.032}$\\
\\
$\omk$&--&&--&--&$0.026^{0.039}_{0.011}$&\\
\\
$n_s$&$0.973^{0.034}_{0.031}$&&$0.974^{0.041}_{0.035}$&$0.959^{0.035}_{0.027}$&$0.968^{0.027}_{0.029}$\\
\\
$log[10^{10} A_s]$&$3.047^{0.135}_{0.083}$&&$3.059^{0.097}_{0.081}$&$3.056^{0.087}_{0.068}$&$3.051^{0.091}_{0.088}$\\
\\
$\w$&$< -0.61$&&$< -0.89$&$< -0.80$&$< -0.77$\\
\\
$\wm$&$ < -0.01$&&$< -0.007$&$0.008$&$< -0.02$\\
\\
$\at$&$< 0.44$&&$< 0.19$&$< 0.33$&$< 0.35$\\
\\
$\dt$&$< 0.37$&&$< 0.21$&$< 0.31$&$< 0.35$\\
\hline
Derived Parameters&&&&&&\\
\\
$H_0$&$59.8^{8.9}_{3.7}$&&$68.9^{4.0}_{3.9}$&$69.2^{2.5}_{3.6}$&$68.2^{5.6}_{4.9}$\\
\\
$\omt$&$0.395^{0.101}_{0.155}$&&$0.285^{0.045}_{0.035}$&$0.289^{0.042}_{0.033}$&$0.295^{0.040}_{0.042}$\\
\\
$\Omega_{DE}/\Omega_m (z_{CMB})$&$ < 0.03$&&$< 0.023$&$< 0.017$&$< 0.025$\\
\\
$\sig$&$0.590^{0.210}_{0.096}$&&$0.713^{0.130}_{0.121}$&$0.806^{0.055}_{0.086}$&$0.724^{0.084}_{0.149}$\\
\end{tabular}\label{tab:cmb}
}
\end{center}
\end{table*}

\end{document}